\documentclass[journal]{IEEEtran}

\usepackage{graphicx}
\usepackage{tabularx}
\usepackage{url}
\usepackage[british]{babel}
\makeatletter
\g@addto@macro{\UrlBreaks}{\UrlOrds}
\usepackage[justification=centering,font={small}]{caption}
\usepackage[noadjust]{cite}
\def\citepunct{,\penalty\citepunctpenalty\hskip.13emplus.1emminus.1em\relax}
\def\citedash{\hbox{--}\penalty\citepunctpenalty}
\begin{document}

\title{Future Scenarios and Challenges for\\Security and Privacy}
 
\author{\IEEEauthorblockN{Meredydd Williams, Louise Axon, Jason R. C. Nurse and Sadie Creese\\}
\IEEEauthorblockA{Department of Computer Science,\\University of Oxford, Oxford, UK\\
\textit{\{firstname.lastname}\}@cs.ox.ac.uk}}

\maketitle

\begin{abstract}
Over the past half-century, technology has evolved beyond our wildest dreams. However, while the benefits of technological growth are undeniable, the nascent Internet did not anticipate the online threats we routinely encounter and the harms which can result. As our world becomes increasingly connected, it is critical we consider what implications current and future technologies have for security and privacy. We approach this challenge by surveying 30 predictions across industry, academia and international organisations to extract a number of common themes. Through this, we distill 10 emerging scenarios and reflect on the impact these might have on a range of stakeholders. Considering gaps in best practice and requirements for further research, we explore how security and privacy might evolve over the next decade. We find that existing guidelines both fail to consider the relationships between stakeholders and do not address the novel risks from wearable devices and insider threats. Our approach rigorously analyses emerging scenarios and suggests future improvements, of crucial importance as we look to pre-empt new technological threats.
\end{abstract} 

\vspace{-0.5em}
\section{Introduction}
\label{sec:one}

Over the past half-century, technology has evolved beyond our wildest dreams. While we once completed our time-shared tasks on chunky mainframes, we now collaborate on documents using our mobile phones. Though these advances have created innumerable benefits for society, they have also brought many risks to security and privacy. Since the Internet was not predicted to face malicious intent, this architecture is now victim to a host of phishing, spam and malware attacks. We currently stand at the dawn of a further technological revolution, with the growth of the Internet-of-Things (IoT) and sophisticated machine learning techniques. As we begin to live our lives `online', it is crucial we consider what future changes might mean for security and privacy.

Due to this criticality, several predictions have been made in industry, academia and international organisations. For instance, The Microsoft Cyberspace2025 \cite{Burt2014} report considers three technological possibilities: a stable but stalled `plateau', a cooperative and innovative `peak', and a `canyon' of fragmented solutions. Choo \cite{Choo2011} approached the topic from a criminological angle, with his scenarios ranging from mobile device malware to highly-sophisticated phishing attacks. The World Economic Forum \cite{WorldEconomicForum2014} collated interviews from global executives and discussed hackers stunting economic growth, threats damaging online services, and resilience triggering further innovation. Through surveying many such scenarios, we found that while reports often differ in scope and audience, there are several common themes.

In this paper we consolidate 30 scenarios from a number of studies to explore those technological futures most frequently considered. Rather than fielding remote predictions, we use a rigorous methodology to identify common existing themes and distil 10 emerging scenarios. We reflect on these possibilities, which include the proliferation of attack tools, before considering their potential impact on a range of stakeholders. We then analyse existing security frameworks and best practices, exploring their suitability in these new environments. Finally, we characterise what research is required to address outstanding security and privacy risks, and consider implications for policy makers and academia.
\section{Methodology}
\label{sec:two}

Our methodology consists of four stages, using an iterative approach to identify those scenarios most-frequently considered. Rather than merely fielding our own predictions, we distil emerging scenarios from a wide range of surveyed literature. These stages are as follows:

\begin{enumerate}
  \item We surveyed a large number of existing security and privacy predictions, many of which are described in Section \ref{sec:three}. These works originated from a range of fields, including academia, industry and international organisations. This was critical to ensure we considered both predictions grounded in technical expertise and those written at a policy level. Regardless of background, these articles were selected based on three key criteria: relevance, citation count and rigour of methodology.  

  \item The works were then indexed and their predictions individually extracted. Whereas several articles clearly delineated between different situations, others required careful inspection to isolate scenarios. We coded these predictions based on their general themes, such as `big data', enabling exploration of which points academia, industry and organisations were in agreement.

  \item Next, we grouped similar predictions, repeating this process in an iterative fashion until we achieved convergence. Where assigned codes concerned related trends, such as `big data' and `machine learning', these were merged to construct rich categories. From our initial 30 scenarios, we distilled 10 predictions which we found most representative. 

  \item Finally, we expanded these scenarios into brief narratives for further analysis. Through considering the themes identified in each instance, we went on to explore how these changes could affect societies commercially, technologically and politically.
\end{enumerate}

\section{Emerging Scenarios}
\label{sec:three} 

We now present our surveyed predictions before consolidating common themes into core emerging scenarios. These works originate from industry, international organisations and academia, providing a comprehensive base for our analyses.

\subsection{Surveyed predictions}

Among industrial predictions, Burt \textit{et al.} \cite{Burt2014} developed the Microsoft Cyberspace2025 report, considering scenarios grounded in an econometric model. In addition to the quantitative analysis of 100 socio-economic indicators, qualitative insights were sourced from researchers and experts. They constructed three future scenarios: `peak' comprising the embracement of technology, `plateau' representing piecemeal cyber adoption, and `canyon' theorising protectionist strategies. Hunter \cite{Hunter2013}, performing industrial research through Gartner, plotted an x-axis of targets against a y-axis of authorities. His scenario quadrants ranged from regulated risk (the establishment of software liabilities) to coalition rule (the growth of underground hackers). In a less formalised approach, ICSPA \cite{InternationalCyberSecurityProtectionAlliance2013} validated their concepts through an expert panel. Through collaboration with the TrendMicro security firm, they discussed a second-generation digital native, a business in a highly-technological market, and a future nation state.

Organisational predictions included that from the World Economic Forum \cite{WorldEconomicForum2014}, who collated interviews and workshops with global executives. Through this, they considered expert feedback before constructing a 14-point roadmap for external collaboration. These findings were drawn into several scenarios, including governments erecting technological barriers and the adoption of new trade innovations. Schwartz \cite{Schwartz2011}, writing for the Center for a New American Security, considered optimistic and pessimistic predictions. Through his analyses of policy and technology he identified both scenarios and indicators of cybersecurity progress. His considerations included the construction of national networks and the proliferation of simple attack tools.

We finally surveyed academic predictions, including scenarios from the Berkeley Center for Long-Term Cybersecurity \cite{BerkeleyCenterforLong-TermCybersecurity2015}. They convened an interdisciplinary conference with experts from computer science and politics. From nascent ideas formed during the sessions, Berkeley graduate researchers developed the concepts into five detailed narratives. These included the Internet becoming an anarchy and the monitoring of human emotions. Benzel \cite{Benzel2015} took an alternative approach, describing discussions from a US Department of Homeland Security session concerning cybersecurity. The event saw a number of working groups deliberating those objectives which should be prioritised over the next five years. Recommendations included increasing the role of the private sector and developing metrics to evaluate cyber investments. In contrast, Choo \cite{Choo2011} conducted an exploration of cybercrime and how this might evolve in the future. Through considering affairs in the US, UK and Australia, he constructed several predictions and suggested criminological prevention strategies. His scenarios ranged from social media propaganda to the escalation of international tensions.

\subsection{Constructed scenarios}

Our 10 emerging scenarios are now discussed in detail, considering the technological, commercial and political possibilities of the next decade. These scenarios are not necessarily complementary, but describe a range of potential eventualities. They are as follows:

\begin{enumerate}
  \item \textbf{Growth of the Internet-of-Things.} The Internet-of-Things (IoT) pervades daily life, blurring the physical and virtual worlds. This entanglement leads to online risks becoming increasingly intangible, contributing to further cyber-threats \cite{Williams2016}. Social norms evolve sufficiently that the IoT is considered `normal' and those who shun these novel technologies are viewed as antiquated. (\cite{Burt2014, Choo2011,WorldEconomicForum2014,InternationalCyberSecurityProtectionAlliance2013,BerkeleyCenterforLong-TermCybersecurity2015})
  \item \textbf{Proliferation of offensive tools.} Offensive government cyber capabilities and the proliferation of simple attack tools both contribute to frequent incidents. Nation-state weaponry leaks onto the black market and is reused by cybercriminals to steal from organisations. Intelligence agencies stockpile zero-day vulnerabilities rather than informing affected vendors, resulting in a deluge of data breaches (\cite{Burt2014,Choo2011,WorldEconomicForum2014,Hunter2013,BerkeleyCenterforLong-TermCybersecurity2015}).
    \item \textbf{Privacy becomes reinterpreted.} The concept of privacy is reinterpreted by `digital natives' who have been raised in an age of social networking and ubiquitous Internet access. Individuals become accustomed to the development of invasive technologies which offer great benefits to convenience and productivity. Lives are lived `online' and reputational damage is frequent as citizens display intimate histories through digital portals (\cite{InternationalCyberSecurityProtectionAlliance2013,BerkeleyCenterforLong-TermCybersecurity2015}).
  \item \textbf{Repressive enforcement of online order.} While many states take liberal risk-based approaches, several favour repressive means to enforce order online. This leads to blanket censorship and surveillance, an order stronger than in current regimes, damaging cross-border trade and placing global commerce under pressure. Although these measures successfully reduce the scale of domestic cyber-attacks, the injury to free enterprise results in economic degradation (\cite{Burt2014,Choo2011,Hunter2013,InternationalCyberSecurityProtectionAlliance2013}).
  \item \textbf{Heterogeneity of state postures.} The heterogeneity of state technology postures stifles international agreement and cooperation over cyber norms. This raises global tensions as attacks increasingly originate from `safe havens' who refuse to prosecute their cyber criminals. The challenge of digital attribution leads to a fragmentation of shared understandings, with nations treating their allies with cool suspicion (\cite{Burt2014,Choo2011,WorldEconomicForum2014,Hunter2013,InternationalCyberSecurityProtectionAlliance2013}). 
  \item \textbf{Traditional business models under pressure.} The concept of intellectual property evolves as traditional business models are placed under increasing pressure from both pirates and new competitors. Established market leaders fall and sell their data assets to innovative firms better-equipped to operate in these novel environments. Cloud platforms become the norm, with individuals granting increased agency to those companies which store their data   (\cite{Burt2014,Hunter2013,InternationalCyberSecurityProtectionAlliance2013,BerkeleyCenterforLong-TermCybersecurity2015}).
  \item \textbf{Big data enables greater control.} Big data and machine learning supports the manipulation of individuals' behaviour by corporations and governments. Sophisticated statistical analysis enables companies to intricately tailor their advertisements, while political parties customise their media campaigns to target individual citizens. While these developments benefit commerce and law enforcement, deviations from the norm are increasingly viewed with suspicion  (\cite{InternationalCyberSecurityProtectionAlliance2013,BerkeleyCenterforLong-TermCybersecurity2015,Benzel2015}).
  \item \textbf{Growth of public-private partnerships.} Organisations continue to know more about their customers than national governments, contributing to the growth of public-private data-sharing partnerships. These arrangements are indeed beneficial to national security, but large parts of critical infrastructure remain owned by foreign corporations. This reliance on private industry shifts considerable power from elected state officials to unaccountable executives, resulting in damage to the democratic process. With power production frequently under the influence of foreign investors, national security becomes critically undermined (\cite{Burt2014,Choo2011,WorldEconomicForum2014,InternationalCyberSecurityProtectionAlliance2013,Benzel2015}).
  \item \textbf{Citizens demand greater control.} Some citizens demand greater transparency and agency over their online data. Technologically-literate individuals store information remotely, occasionally selling details for a range of benefits. Corporations offer paid alternatives to data-hungry social networks, creating new markets for online communities and Privacy-Enhancing Technologies (PETs) (\cite{Choo2011,Hunter2013,InternationalCyberSecurityProtectionAlliance2013,BerkeleyCenterforLong-TermCybersecurity2015}).
  \item \textbf{Organisations value cyber-resilience.} Cyber-resilience is increasingly important for informing business decisions, with issues such as insider threats high on the agenda. The infeasibility of absolute security drives a market for cyber-insurance, as corporations adopt prescribed measures to reduce their premiums. Board rooms quickly learn the reputational costs of cyber-attacks, resulting in greater private investment in the actuarial and mathematical sciences (\cite{Choo2011,WorldEconomicForum2014,Hunter2013,InternationalCyberSecurityProtectionAlliance2013,Benzel2015}). 
\end{enumerate}

\vspace{-1.0em}

\subsection{Stakeholder impact}

We continue by considering our emerging scenarios and the impact they could have at three stakeholder levels: \textit{individual}, \textit{organisational} and \textit{national}. Investigation of stakeholder impact is crucial to explore how ordinary citizens, corporations and governments may operate in the future. Considerations are summarised below in Fig. \ref{fig:one}, with yellow, orange and red representing low, medium and high impact, respectively. These ratings were determined through identifying which scenarios could critically affect a stakeholder, which would require changes in practice, and which would have a limited impact.

\begin{figure}[h!]
    \includegraphics[width=0.47\textwidth]{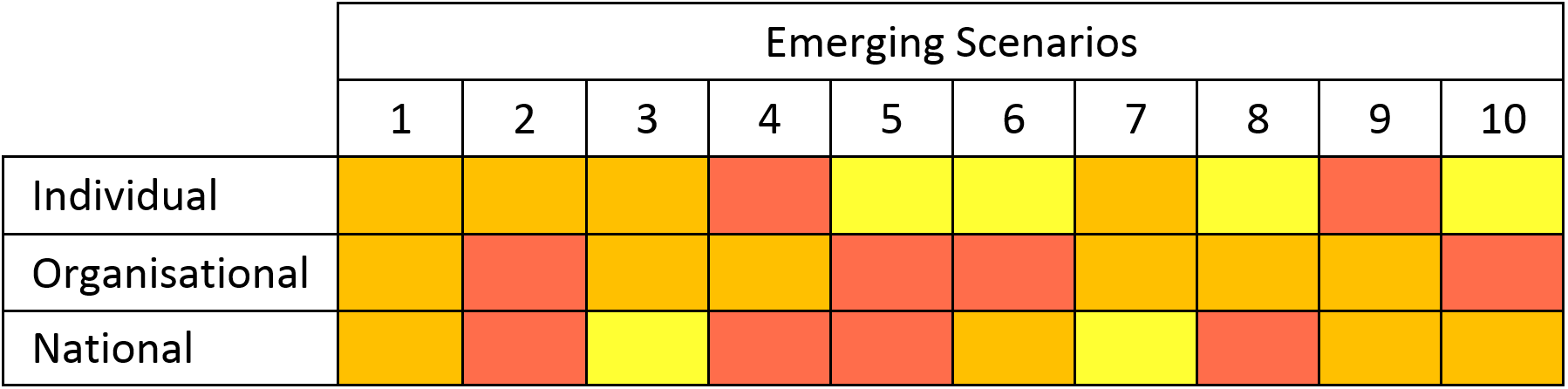}
    \centering
    \caption{Emerging scenarios and stakeholder impact}
    \label{fig:one}
\end{figure}

In detail, the stakeholder impacts are as follows:
\begin{enumerate}
  \item The pervasion of the \textbf{Internet-of-Things} would present dangerous opportunities for the surveillance of individuals, particularly if biological data is accessible through implanted devices. While organisations would benefit from enhanced home-working efficiencies, they must defend against Bring-Your-Own-Device (BYOD) risks. Although national governments would benefit economically from the IoT, the dependence of critical infrastructure on modern technologies would greatly increase attack surfaces.
  \item \textbf{The proliferation of attack tools} would place individuals under the ubiquitous risk of identity theft. Organisations similarly must defend against data exfiltration (of high impact) and therefore might be less willing to store customer data. International tensions would be fraught as high-profile attacks become commonplace, especially with attribution appearing largely intractable.
  \item \textbf{Reinterpretations of privacy} would see PII (Personally Identifiable Information) lose its importance as more-sensitive details, such as sexual orientation, are available through social media. Although organisations encourage the exchange of data for convenience, they are at risk of their own employees sharing confidential information. Nations benefit from large collections of open source intelligence, but find their citizenry increasingly dependent on services hosted in foreign countries.
  \item \textbf{The repressive enforcement of online order} would involve widespread filtering, censorship and surveillance, highlighted as high impact in Fig. \ref{fig:one}. Organisational opportunities would be stifled by both technological restrictions and costly regulatory compliance. Foreign opposition to repressive policies would damage international trade, while prosecuting dissidents would be both challenging and expensive.
  \item \textbf{The heterogeneity of state postures} would result in minimal cross-border personal data protection, especially when nations differ in their definitions of PII. Unpredictable international relations would contribute to market instability, while high-impact cyber attacks continue from `safe havens'. Reductions in information sharing would damage law enforcement, with global tensions increasing through frequent disagreements.
  \item \textbf{The evolution of new business models} would see individuals' personal data become the most valuable commodity, frequently traded on global exchanges. Increased market competitivity would depress profit margins while organisations reel from ubiquitous copyright infringement. National economies are similarly weakened by intellectual property violations and lose stability as established firms are replaced by start-ups. With citizen data increasingly residing in foreign data centres, both individuals and governments would be vulnerable to the decisions of other states.
  \item \textbf{Advances in big data and machine learning} would result in consumer manipulation and reductions to their individual agency. Organisations would benefit from sophisticated financial forecasting and seamless targeted advertising. Nations better control their citizenry through advanced media management, yet increasingly rely on a technology bubble for their economic growth.
  \item \textbf{Public-private data-sharing partnerships} would infringe individuals' rights as their personal data is shared with foreign governments. The wider distribution of this information would also increase the risk of confidentiality being breached. Although organisations might learn valuable threat intelligence, they could have their reputations damaged by governmental arrangements. While national agencies would acquire valuable data, foreign states might exert an troubling corporate influence over critical infrastructure. 
  \item Although \textbf{citizen demands} might give certain individuals greater freedom, technical literacy is required and commercial costs would be inherited through price increases. Organisations struggle under the pressure to reveal corporate information while national governments find themselves increasingly accountable to their citizens. Despite both initiatives being beneficial to democracy, they weaken domestic economies against unaccountable rival states.
  \item \textbf{The importance of cyber-resilience} emphasises the ubiquitous personal risk from data breaches. Organisations would invest in costly technological protection, but still fall victim to high-impact Advanced Persistent Threats (APTs). Insider threats would challenge the sensitive world of government, particularly since the capabilities of malicious employees will increase with ubiquitous  technology. As a result, frequent breaches in both the public and private sectors could undermine trust in state institutions.
\end{enumerate}

\section{Challenges for Security and Privacy}
\label{sec:four} 

As a next stage, we reflect on the future applicability of current `best practice' guidelines for \textbf{individuals}, \textbf{organisations} and \textbf{nations}. This is of great importance, as these documents inform stakeholder groups and are expected to guard against technological threats. We summarise the gaps in current frameworks, selecting several key examples due to space restrictions, and point to required research and development. At the individual level, we consider the main UK and US government initiatives for improving cybersecurity practices: for the UK, the Cyber Streetwise campaign \cite{cyberstreetwise}, and for the US, Stop.Think.Connect \cite{stopthinkconnect}. The UK Get Safe Online initiative was also considered, but this effort is private sector and less authoritative than Cyber Streetwise. For organisations, we examine the US National Institute for Standards and Technology's (NIST) cybersecurity framework \cite{NIST}, and the Center for Internet Security's Critical Security Controls (CIS CSC) \cite{sans}. The International Telecommunication Union (ITU) National Cybersecurity Strategy Guide \cite{ITU} and NATO National Cybersecurity Framework Manual \cite{NATO} are considered as internationally-recognised guidelines for nations. Based on our findings, we conclude this section with a call to action for researchers, stakeholders and policy makers.

\subsection{Gaps in existing guidelines}

While advantageous for the present day, \textbf{awareness campaigns for individuals} do not translate to a future in which the IoT pervades daily life. In this future, individuals' understanding of, and agency over, their online presence is key to the protection of their privacy. Current campaigns go some way to promoting good practice in this area (e.g. recommendations for protecting privacy when using social media \cite{cyberstreetwise}; privacy-protecting recommendations given for the use of web services, devices, purchase history, and location \cite{stopthinkconnect}). However, these initiatives do not extend to educating citizens on the privacy implications of increasingly-used wearable, biometric and sensor-based technologies. Nor is clear usable advice given on the specific types of data collected by different applications and devices. For guidelines to be suitable for future scenarios, campaigns must focus on emerging technologies and give specific, practical advice. Furthermore, these public initiatives should be periodically evaluated to ascertain their effectiveness in improving individuals' behaviour. Whether undertaken through large-scale surveys or targeted focus groups, feedback mechanisms would enable campaigns to be iteratively refined. 

\textbf{Existing frameworks for organisations} do not adequately address those human aspects which will be key to future corporate cybersecurity. The NIST and CSC documents \cite{NIST, sans} are highly comprehensive, with useful references to external guidelines. They do not, however, provide frameworks within which the human aspects of organisational cybersecurity can be addressed; namely, problems arising from insider threats and BYOD. For instance, NIST aims to mitigate insider threats by recommending the comparison of personnel actions against expected behaviours and data-flow baselines. CSC addresses the problem by advocating the controlled use of administrator privileges, access controls on a ``need to know'' basis, and the monitoring of employee and contractor accounts. While these activities are essential, employee tracking alone will not defend against insiders who are aware of the monitoring systems in place \cite{colwill2009human}. Unfortunately, both frameworks fail to give any consideration to the growing need to study personnel motives or to monitor behaviour from a psychological perspective.

Both the NIST and CSC recommendations protect organisations against BYOD risks insofar as they specify that physical devices should be inventoried and monitored. We foresee a huge expansion in the ownership of IoT products, resulting in an increase in the number of personally-owned devices present in the workplace; in particular, wearables and implants. In this future, registering all connected products will not be feasible. Sophisticated means of mitigating IoT and BYOD risk are required \cite{Nurse2015}, such as protection against potentially-invisible devices (e.g. implanted medical chips) which are challenging to catalogue. This is crucial to recognising the increasing corporate blur between internal and external spaces.

\textbf{Cybersecurity frameworks for nations} \cite{ITU, NATO} recommend international cooperation to unite national strategies. Set against our future scenarios, this is extremely optimistic, especially between non-allied states. Indeed, it is predicted that heterogeneous state postures and differing cybersecurity laws will stifle international cooperation. This combined with the proliferation of cyber weaponry and problems of attribution suggests that the alignment of national strategies is unrealistic. Of greater importance is the construction of policies and legislation which allows states to suit their own, and international, interests in a non-unified cyber landscape. 

Based on our scenarios, we foresee that both state cyber weaponry will grow and international tensions will escalate as a result of high-profile incidents. A recent example of expected events is the unattributed Ukrainian power outage of December 2015, apparently caused by a cyber-attack \cite{BBCNews2016}. Currently, agreement of those terms key to building legislation is lacking. For example, there are no widely-agreed definitions for either ``cyber warfare'' or ``cyber attack'' \cite{NATO}. Legislation in this area is crucial to maintaining order in the coming decades. 

\subsection{Requirements for research and development}

Based on the gaps identified, much work is required to extend cybersecurity frameworks for the future. We discuss areas that must be addressed through academic research, the development of policy, and action by stakeholders.

\textbf{For the protection of individuals' privacy}, policies and laws on collection, storage and processing must be reconsidered in light of the increased sensitivity and abundance of data. Education will be central to afford citizens the understanding to retain agency over their information, with progress monitored through national surveys. The development of disclosure metrics for individuals and organisations will help technology users perceive the risks of their decisions, and assist regulators in detecting abuse. Similarly, methods to predict the lifetime and impact of sensitive data would enable individuals to understand their risk exposure. Usable privacy solutions should also be developed, such as applications which highlight the data collected by wearable devices.

\textbf{For organisations}, research into the human aspects of insider threat, such as behavioural psychology and criminology, is important. Both the corporate adoption of education schemes and the consideration of employees' actions are critical to proactively mitigating these risks. Whilst existing studies have began to explore the psychological, behavioural, cultural and technical characteristics of insider threats \cite{Nurse2014}, further research is required to develop comprehensive solutions that are ultimately useful for organisations. The evolution of current practice is crucial if companies wish to maintain resistance against increasingly-sophisticated insider threats. 

BYOD currently poses a number of organisational challenges \cite{Nurse2015,garba2015review} and our emerging scenarios suggest this situation will only become worse. To protect companies against the increased number and variety of personal devices brought to the office, two developments are required. Firstly, applications should be built which secure company networks against personal technologies. Secondly, effective organisational policies should be implemented to manage risk in a future pervaded by wearable and implanted devices.

\textbf{At the national level}, more specificity is required in calls for international cooperation. Whilst some states will negotiate mutual pledges, as the US and China did in their 2015 Cyber Agreement \cite{Rosenfeld2015}, the global unification of security postures is highly unlikely. As such, both international cybersecurity bodies and frameworks for addressing key matters, such as piracy and cross-border law enforcement, must function in a heterogeneous environment. We predict the ownership of large parts of critical national infrastructure by foreign corporations, granting power to foreign investors and underlying states. In a cyber landscape in which nations are not concordant, legal agreements and technological solutions are required to address external control over critical infrastructure.

As offensive cyber weapons proliferate and global tensions increase, legislation on `acts of war' and digitally-proportionate responses must be defined. It is inevitable that nation-state cyber weaponry will reach the black market, and therefore it is crucial the consequences of its use are understood \cite{hodgescreese}. Furthermore, strong controls should be enforced to avoid the escalation of these most harmful risks. 

\textbf{Interactions between individuals, organisations and nations} must also be considered. While we have explored the future scenario impact on our three stakeholders individually, in reality, there are several relationships and dependencies between these parties. For example, the effect that advances in machine learning will have on individuals' privacy will depend largely on both the regulations governing data collection and the behaviour of organisations. There is, however, no single document that details cybersecurity guidelines across these three differing levels. This lack of cohesion introduces complex subtleties and key dependencies can be missed. Broad frameworks would enable a clear vision of both current cybersecurity practices and their future weaknesses. However, it might be challenging to locate a party capable of constructing such a comprehensive document. This is especially true considering the international scope of the field, with recommendations required to be equally applicable to a Chinese power station and a US bank. Specificity is naturally in tension with generality, and new guidelines would also require frequent update to ensure they are suitable. These factors, and the difficulty of selecting an authorship trusted by all international parties, ensure that the construction of a comprehensive document would be challenging. However, such efforts are essential if we wish to protect security and privacy in an uncertain global future.

Perhaps a more achievable proposal would be to increase collaboration between guideline authors at the individual, organisational and national levels. Such unification of approaches and ideas would enable the production of frameworks which are more strongly contextualised against cybersecurity at all levels. We illustrate this point by drawing on our above example, which is based on advances in machine learning and consumer prediction capabilities. This scenario highlights a discrepancy between the recommendations made at the organisational and individual level. Given that guidelines in the former \cite{NIST, sans} provide only limited advice on the use of sensitive consumer information, the latter's \cite{cyberstreetwise, stopthinkconnect} recommendations on data protection is inadequate to maintain privacy in a world of sensor-based devices. In the space between these two frameworks, there are real risks of organisational data use against individuals' wishes. These gaps could be bridged by collaboration between the authors of such guidelines.  

\section{Conclusions}
\label{sec:five}

Emerging scenarios provide useful insights into the future of security and privacy. While previous predictions are diverse in their scopes and audiences, through constructing representative futures, we explored those developments most frequently-anticipated. Our study of the impact on individuals, organisations and nations highlights key considerations for these important stakeholders. We also analysed existing frameworks and best practices, identifying guidelines which do not adequately translate to future predictions. Our findings suggest avenues for research and development, including privacy education for the general public, approaches to mitigate organisational insider threat, greater investigation of BYOD risk, and international legislation for cyber warfare. It is crucial that emerging scenarios are explored in anticipation of the security and privacy threats of the coming decade.

We have considered a number of possibilities for future work. Firstly, we could conduct a technical analysis of why reputable guidelines fail to address the threats from emerging scenarios. In this work, we could rigorously examine individual sections in `best practice' documents and identify which areas urgently require updating. Secondly, these emerging scenarios could be explored within the context of specific domains, such as healthcare. The pervasion of the Internet-of-Things might be beneficial with the development of wireless sensors, but also threaten welfare when critical devices are compromised. Similarly, although advances in big data might revolutionise the detection of health conditions, these measures could unfairly increase insurance premiums for victims of profiling. Finally, we could expand our scope and explore future possibilities for technology as a whole. The coming decades promise to radically change how we perceive digital devices, and it is crucial we are prepared for these novel developments.

\bibliographystyle{IEEEtran}
\bibliography{bib}

\end{document}